\documentclass[prl,preprint,double-spaced,showpacs,preprintnumbers,amsmath,amssymb]{revtex4}
\usepackage{syntonly}
\usepackage{textcomp}
\usepackage{hyperref}
\usepackage{graphicx}
\usepackage{amssymb}
\begin{document}
\title{Evidence for electronic phase separation between orbital orderings in SmVO$_3$}
\author{M.H. Sage$^1$, G.R. Blake$^1$, G.J. Nieuwenhuys$^2$, and T.T.M. Palstra$^1$}
\altaffiliation [Corresponding author, ]{email:
t.t.m.palstra@rug.nl}
\affiliation{$^1$Solid State Chemistry Laboratory, Materials Science Centre,\\
University of Groningen, Nijenborgh 4, 9747 AG Groningen, The
Netherlands. \\$^2$Kamerlingh Onnes Laboratorium, Leiden
University, P.O. Box 9504, 2300 RA Leiden, The Netherlands.}

\begin{abstract}
We report evidence for phase coexistence of orbital orderings of
different symmetry in SmVO$_3$ by high resolution X-Ray powder
diffraction. The phase coexistence is triggered by an
antiferromagnetic ordering of the vanadium spins near 130K, below
an initial orbital ordering near 200K. The phase coexistence is
the result of the intermediate ionic size of samarium coupled to
exchange striction at the vanadium spin ordering.

\end{abstract}


\maketitle Transition metal oxides with competing interactions
have in recent years been shown to exhibit complex electronic
behavior, in which different electronic phases may coexist. The
competing interactions include the spin, charge and orbital
degrees of freedom. The origin of the phase coexistence is often
associated with a coupling of the electronic degrees of freedom to
the lattice. Much interest has been triggered by colossal
magneto-resistance materials where a metal-insulator transition
involves a discontinuous change in the molar volume
\cite{112,260,261,262}. Such a first order transition can lead to
phase coexistence of metallic and insulating states, where the
length scales of these states are determined by strain and stress.
In (La,Sr)MnO$_3$ and (La,Ca)MnO$_3$ \cite{124} such phase
coexistence has been observed for low concentrations of the
divalent cation as droplets, and it results from competition
between the 'potential' and 'kinetic' energy of the valence
electron system. Electron delocalization is promoted by kinetic
energy, whereas localization is promoted by the Coulomb repulsion
between electrons. In manganite systems extremely diverse types of
phase separation can develop, occurring on a variety of
length-scales that can range from micrometers down to a few
nanometers. Interestingly, phase coexistence not associated with
metal-insulator transitions has also been observed; it can also
arise from competition between different charge- and
orbitally-ordered or disordered phases \cite{268,269}. The spin,
charge and orbital orderings found in the manganites are often
associated with large displacements of the ions in the lattice or
even with changes in the molar volume. Generally, the orbital and
charge ordering take place at higher temperatures than the
magnetic ordering \cite{148}. In both cases the coupling between
the different phases is mediated through coupling with the
lattice. In contrast, the coupling of orbital and spin-order is
mostly of electronic nature by the antisymmetrization requirement
of the wavefunction. In LaMnO$_3$ \cite{263} the orbital
interactions between the e$_g$ electrons are much stronger
(T$_{o.o}$ $\sim$ 800K) than the magnetic interactions (T$_N$
$\sim$ 150K). In contrast, the Jahn-Teller active t$_{2g}$
electrons in the vanadates orbitally order at T$_{o.o}$ $\sim$ 200
K or lower. Although the difference between the spin and orbital
ordering temperatures in the vanadates is smaller than that in the
manganites, relatively little interaction between the spin and
orbitally ordered states is observed. The orbital ordering
temperature is higher than the magnetic ordering with the
exception of LaVO$_3$. For the small ionic size R in RVO$_3$,
complex behavior has been observed in which different types of
orbital ordering occur \cite{2,202}. Here, a change in the orbital
structure is accompanied by a change in spin order. This indicates
a strong coupling between the two types of order as required by
the antisymmetric nature of the electron wavefunction, including
spin and orbit. Nevertheless, the orders originate at distinctly
different temperatures. Here, we report the orbital ordering of
SmVO$_3$, in which coexistence of orbital orderings of different
symmetry is triggered by magnetic ordering of the vanadium spins.
This coexistence of orbital orderings is not associated with
metal-insulator transitions, but is the result of magnetic
exchange striction, which provides the coupling to the lattice.
\par Polycrystalline powder samples of SmVO$_3$ were prepared by
chemical reduction of SmVO$_4$ at 1400\textdegree C under a
H$_2$/N$_2$ atmosphere. The SmVO$_4$ powders were prepared by
solid state reaction, using pre-dried Sm$_2$O$_3$ (99.9\%), and a
10\% excess of V$_2$O$_5$ (99.95\%) to compensate the
V-volatility. The sample shows a small impurity of V$_2$O$_3$ due
to the large excess of V$_2$O$_5$ in the starting material. The
high resolution X-Ray powder diffraction experiments were carried
out on beamline ID31 at ESRF in Grenoble. During cooling down to
5K, short scans were performed. The temperature was then
stabilized at 5K, and a long scan was measured. Then, short scans
up to 254K were measured, while heating to room temperature (RT).
The powder diffraction patterns were analyzed by the Rietveld
method \cite{162} using the refinement program GSAS \cite{253} as
implemented in the EXPGUI package \cite{254}. The specific heat
was measured between 3 and 300K using a relaxation method in a
commercial Quantum Design Physical Properties Measurement System.
\par
\begin{figure}
  \includegraphics[width=8cm]{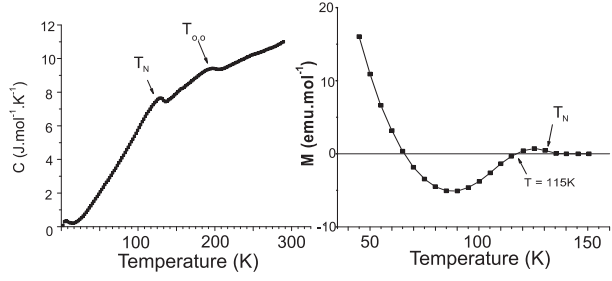}\\
  \caption{Specific heat of SmVO$_3$ (left) and temperature dependence of magnetization (measured in 1kOe) (right).}\label{HC}
\end{figure}

\begin{figure}
  \includegraphics[width=8cm]{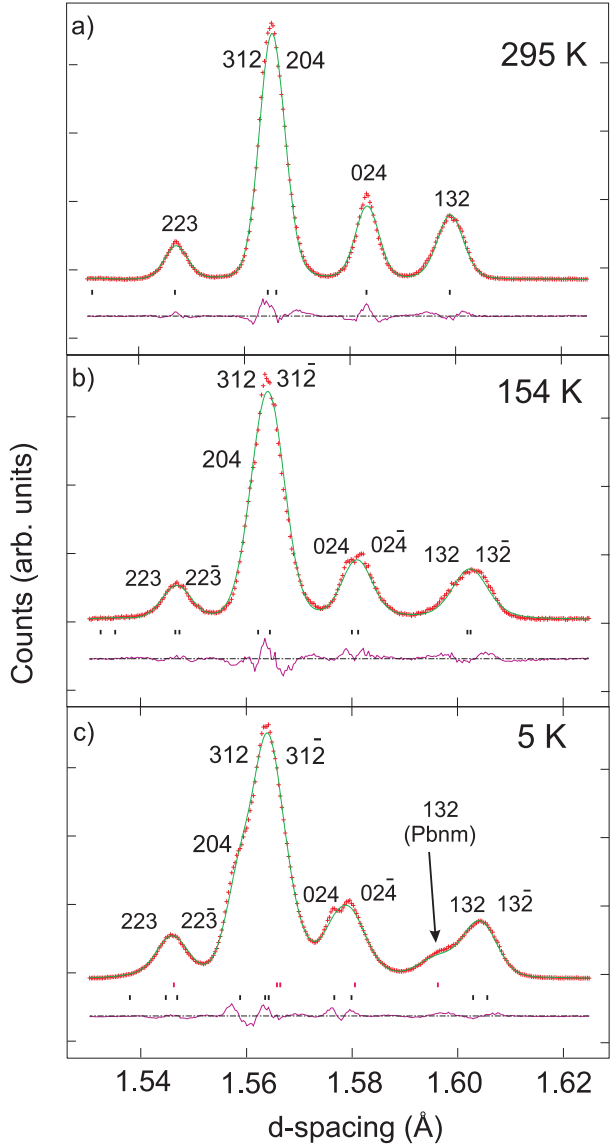}\\
  \caption{Section of the diffraction pattern that shows the successive phases
 observed for SmVO$_3$ with a): T$>$T$_{o.o}$: Pbnm (R$_{wp}$=0.078),
b): T$_{N}$$<T<$T$_{o.o}$: P2$_1$/b11 phase, c): T$<$T$_{N}$: Pbnm
(top markers) and P2$_1$/b11 (bottom markers) phases
(R$_{wp}$=0.0517).}\label{peak}
\end{figure}

 SmVO$_3$ has an orthorhombic perovskite structure at room
temperature, as is common for all RVO$_3$, with R a rare earth
element or Yttrium. The structure consists of corner sharing
oxygen octahedra. The vanadium is located in the center of the
octahedron, and the samarium is located between the octahedra. We
could index the peaks belonging to the SmVO$_3$ phase with the
orthorhombic Pbnm space group. From the refinement of the
diffraction pattern, we observed that SmVO$_3$ stabilizes in an
O-Orthorhombic structure at room temperature \cite{267}
($a<c/\sqrt{2}<b$) because of the buckling of the corner-shared
octahedra \cite{197}. SmVO$_3$ retains its orthorhombic structure
down to T$_{o.o}$=200K. Here the symmetry changes to monoclinic
P2$_1$/b11\footnote{We use P2$_1$/b11 for the monoclinic phase
instead of the standard setting P2$_1$/c in order to allow a
direct comparison of the lattice parameters, Miller indices and
atomic coordinates of the monoclinic phase with those of the
orthorhombic Pbnm phases.} as evidenced by the splitting and
broadening of the diffraction lines. We identify this change in
symmetry with the orbital ordering transition. This is confirmed
by the specific heat measurement, as shown in Fig.\ref{HC}, and is
consistent with other reports \cite{202}.
\begin{figure}
  \includegraphics[width=8cm]{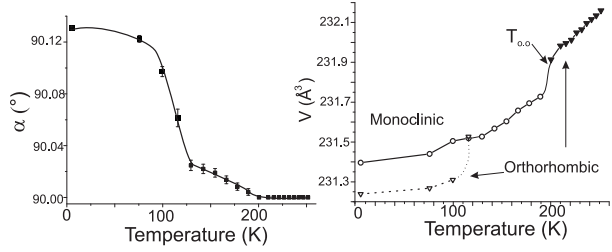}\\
  \caption{Temperature dependence of the alpha angle (left), and of the
 unit-cell volume (right) in SmVO$_3$.}\label{alpha}
\end{figure}
The value of the monoclinic angle increases slowly from
$90.00\,^{\circ}$ at 200K down to 9$0.019(4)\,^{\circ}$ at 129K
(Fig.\ref{alpha}). As the temperature further decreases, the value
of alpha increases rapidly to a maximum value of
$90.131(1)\,^{\circ}$ below 50K (Fig.\ref{HC}). The onset of
orbital ordering is accompanied by a change in the unit-cell
volume (Fig.\ref{alpha}). This monoclinic phase persists from the
T$_{o.o}$ down to the magnetic transition at T$_N$
(Fig.\ref{peak}(b)), which is observed by both specific heat and
magnetization measurements (Fig.\ref{HC}).
\\Below 115K we observe a coexistence of phases with monoclinic
(P2$_1$/b11) and orthorhombic (Pbnm) symmetries, which remains
down to our lowest measurement temperature of 5K
(Fig.\ref{peak}(c)). The proportion of orthorhombic phase
increases from 15\% at 115K to 25\% at 5K. Similar ratios were
observed both upon cooling and warming the sample. We note that
the phase coexistence sets in continuously below 115K, where both
phases have the same molar volume.
\begin{table}{\small
\begin{tabular}{|c|c|c|c|}
  \hline
   & 5K (Pbnm) & 5K (P2$_1$/b11) & 295K (Pbnm)\\
  \hline
  V1-O1& 2.026(5) & 1.996(15) & 1.9942(12)\\
  V1-O2& 1.977(12) & 1.973(16) & 2.0236(35)\\
  V1-O2& 2.025(12) & 2.070(14) & 2.008(34)\\
  V2-O1& & 1.934(15) & \\
  V2-O3& & 2.024(13) & \\
  V2-O3& & 2.049(16) & \\
  \hline
\end{tabular}}
\caption{V-O bondlengths ($\AA$) for SmVO$_3$ in the orbitally
ordered phases at 5K and non-ordered phase at 295K.}\label{bond}
\end{table}
\begin{figure}
  \includegraphics[width=8cm]{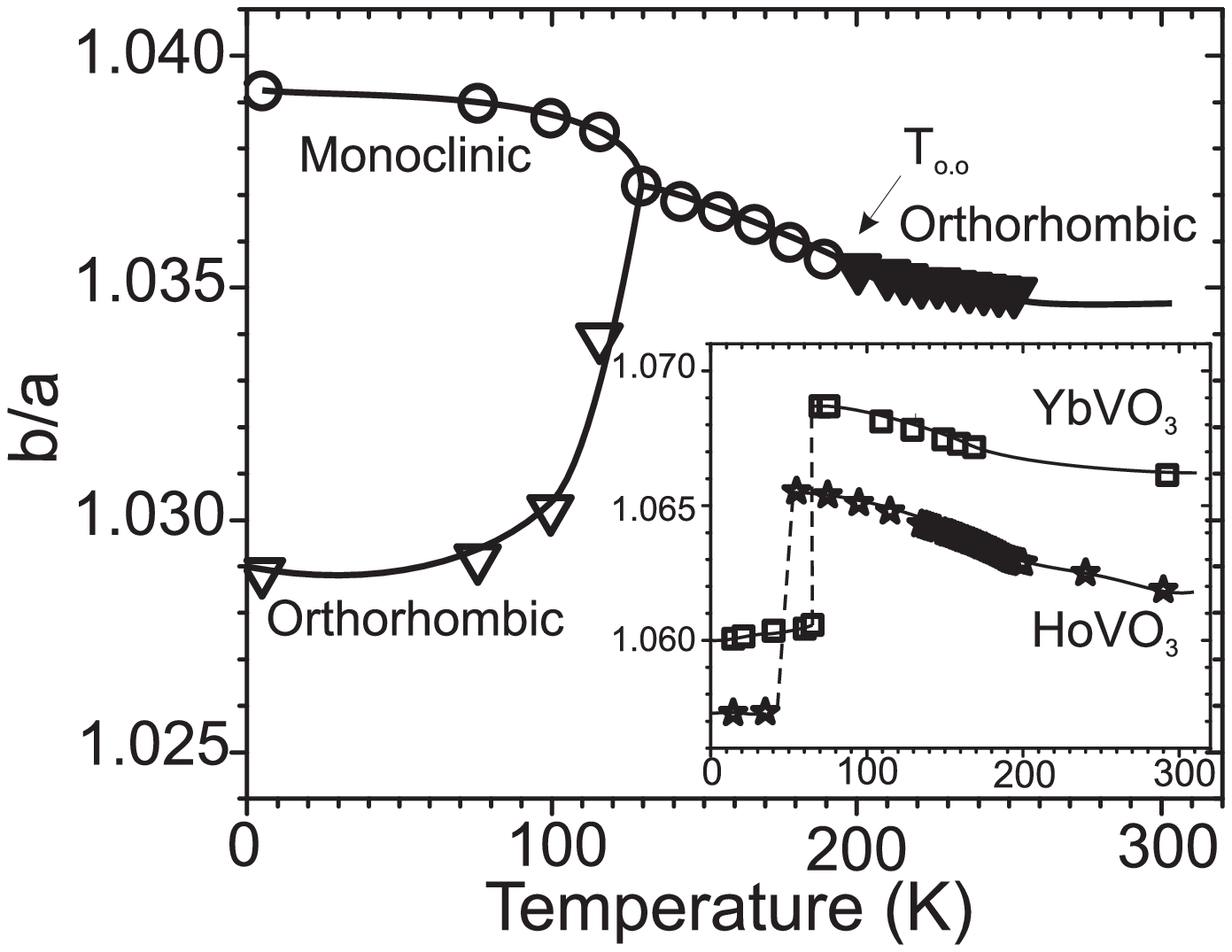}\\
  \caption{Temperature dependence of the lattice parameter ratio b/a
in SmVO$_3$: the filled triangles represent the orthorhombic phase
above T$_{o.o}$). The inset presents this ratio for other RVO$_3$
with R=Ho (stars) and Yb (squares).}\label{param}
\end{figure}
The V-O1, V-O2 and V-O3 distances, where O1 are the out-of plane
and O2 and O3 are the in-plane oxygens, are important indicators
of the type of orbital ordering of a system. In our data they are
difficult to determine accurately because Sm, a heavy rare-earth,
dominates the diffraction pattern and also because there is
considerable overlap of peaks from the coexisting phases at low
temperatures. However, as shown in Table \ref{bond}, the V-O
distances confirm that the RT structure is not orbitally ordered
and strongly suggest that at 5K both the monoclinic and
orthorhombic phases are orbitally ordered. The V-O distances in
the monoclinic and orthorhombic phases are comparable within error
bars to those in the so-called G-type and C-type orbitally ordered
phases of YVO$_3$, respectively \cite{4}. A full list of refined
atomic parameters, bond lengths and angles is deposited in the
auxiliary material. To confirm the nature of these phases over the
whole temperature range, since the rapidly collected
cooling/heating data do not allow a precise determination of the
oxygen fractional coordinates\footnote{These datasets were
collected over 5-10 minutes and, although sufficient for a precise
determination of the lattice parameters, cell volume, monoclinic
angle and phase fractions, below T$_{OO}$ the accurate
determination of oxygen fractional coordinates was not possible;
the atomic parameters were thus fixed to those determined for the
P2$_1$/b11 and Pbnm phases at 5K.}, we have chosen the lattice
parameter ratio b/a (Fig.\ref{param}) as an indicator of the type
of orbital ordering and compared it to the ratios for other
RVO$_3$ \cite{258}. This comparison provides further evidence that
the low temperature orthorhombic phase of SmVO$_3$ is also
orbitally ordered. For YVO$_3$ \cite{4}, HoVO$_3$ and YbVO$_3$,
the transition from G- to C-type orbital ordering is discontinuous
first order and a large drop of the b/a ratio can be observed.
Moreover, it occurs at much lower temperature than the magnetic
ordering. In contrast, for SmVO$_3$ the transition is continuous
and is observed to set in very close to T$_N$.
\par The type of orbital ordering (OO) present in the ground state
of the RVO$_3$ compounds depends mainly on the degree of
octahedral tilting caused by the deviation in size of the
rare-earth cation from that in the ideal cubic perovskite.
Although all RVO$_3$ compounds initially order in the so-called
G-type structure (d$_{xy}$d$_{yz}$ and d$_{xy}$d$_{xz}$ orbitals
are occupied alternately along all three directions of the
lattice), on increasing the octahedral tilting as the rare-earth
cation becomes smaller a first-order transition to a C-type OO
ground state (d$_{xy}$d$_{yz}$ and d$_{xy}$d$_{xz}$ orbitals are
occupied alternately in the $ab$ plane while the same orbitals are
occupied along $c$) occurs in the range 50 K to 80 K for
rare-earths smaller than Tb \cite{202}. This phase transition
involves a large decrease in unit cell volume. In these strongly
distorted structures, the C-type ground state is largely
stabilized by a shift of the R cation in order to maximize the R-O
covalency \cite{264,2}.
\par SmVO$_3$ lies far into the G-type OO
region of the phase diagram. As observed in both our specific heat
and diffraction measurements, G-type OO takes place in SmVO$_3$ at
200 K and the resulting monoclinic distortion becomes
progressively larger with decreasing temperature, as indicated by
the monoclinic angle, alpha, shown in Fig.\ref{alpha}. However,
below ~105 K the value of alpha shows a rapid increase before
reaching a constant value below 50 K. More surprisingly, ~15\% of
the sample is transformed into the C-type OO structure in the
vicinity of T$_N$. This C-type phase persists on further cooling
with a fraction that gradually increases to ~25\% at 5 K. Both the
appearance of the C-type phase and the rapid increase in the
monoclinic distortion of the G-type phase are most likely
associated with the onset of magnetic ordering.
\par It has long been
known that a structural distortion, known as magnetic exchange
striction, can occur at or shortly below T$_N$ in transition metal
oxides in order to increase the magnetic exchange interaction
energy \cite{265}. This type of distortion has previously been
observed in LaVO$_3$ \cite{55,1} and CeVO$_3$ \cite{55}. In the
case of CeVO$_3$ the type of OO remains the same, but the exchange
striction occurring at T$_N$ involves a small increase in a and b,
a large decrease in c, and an overall decrease in the unit cell
volume. In the case of LaVO$_3$, G-type OO arises 2 K below the
magnetic ordering temperature and may itself be the result of
exchange striction. However, in the "small radius" RVO$_3$
compounds such as YVO$_3$, the structural change at T$_N$ is tiny
\cite{247}. The difference between the orbital and magnetic
ordering temperatures in these compounds is much greater than in
LaVO$_3$ and CeVO$_3$ \cite{202} and V-O bond lengths suggest that
orbital ordering is fully developed at T$_N$ \cite{2}. Therefore,
there is little further energy to be gained through a structural
distortion.
\par The exchange striction occurring at T$_N$ in SmVO$_3$
is expected to be intermediate in magnitude between that in
CeVO$_3$ and YVO$_3$; the large increase in the monoclinic angle
alpha below T$_N$ (Fig.\ref{HC}) suggests that orbital ordering in
the G-type phase is incomplete at this point. Since the exchange
striction appears to cause a volume decrease, one may expect the
onset of magnetic ordering to promote the C-type phase; that is,
the C-type phase will be lowered in energy to an extent that
depends on the magnitude of the exchange striction. For compounds
such as CeVO$_3$ the G-type phase is much lower in energy at all
temperatures due to the smaller degree of octahedral tilting. At
the other end of the phase diagram the exchange striction in the
G-type phase at T$_N$ is too weak to have any effect on the
crystal structure. However, for materials closer to the phase
boundary such as SmVO$_3$, the exchange striction is still
significant and may be strong enough to lower the energy of the
C-type phase enough to become favored.
\par Somewhat surprisingly, in
SmVO$_3$ the unit cell volumes of the two OO phases are equal (at
the precision of our diffraction measurements) immediately below
T$_N$ (Fig.\ref{alpha}). However, this can be explained by a
scenario where small droplets of the C-type phase are initially
formed in a G-type matrix. Although the C-type droplets can be
distinguished by X-ray diffraction, implying coherent OO over
length-scales of at least several hundred angstroms, they are
still relatively small and isolated compared to the surrounding
G-type phase. The unit cell volume of regions of a droplet close
to an interface will be forced to match that of the surrounding
matrix and large strain fields will thus extend throughout
droplets that are small. On further cooling, the C-type droplets
will tend to enlarge and coalesce into domains big enough for
relaxation of the strain to take place. This relaxation is seen in
the rapid decrease of the unit cell volume as the fraction of the
C-type phase grows. However, the remaining strain might still be
large enough to inhibit further transformation of the G-type
matrix, stabilizing the phase-separated state with little change
in phase fractions below ~80 K. The difference in unit cell
volumes of the two phases at 5 K is 0.067\%, which can be compared
to changes of 0.16\% on cooling through the first-order transition
in YVO$_3$ \cite{4}, 0.15\% in HoVO$_3$ and 0.16\% in YbVO$_3$
\cite{258}, where full transformation to the C-type phase is
achieved. The suppressed volume decrease in SmVO$_3$ is further
evidence that significant strain remains in the system down to 5
K.
\begin{figure}
  \includegraphics[width=8cm]{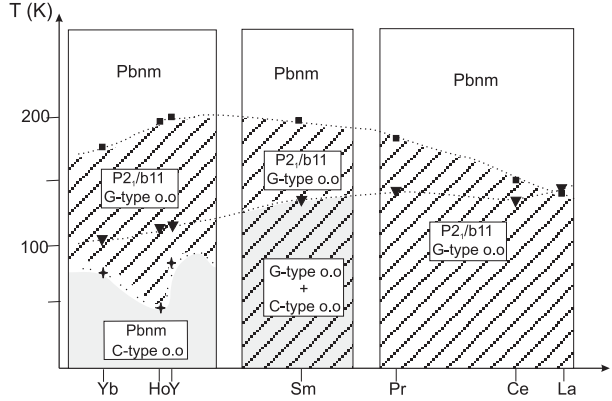}
  \caption{Phase diagram obtained for RVO$_3$ studied by diffraction techniques.
The squares represent the onset of orbital ordering, the triangles
represent spin ordering, stars represent the first order
transition.}\label{phase}
\end{figure}
\par Although specific heat measurements show no transition between
orbital orderings below T$_N$ in other "intermediate radius"
vanadates such as EuVO$_3$, GdVO$_3$ and TbVO$_3$ \cite{202}, the
low-temperature phase composition of these materials has thus far
not been studied by techniques such as high resolution X-ray
diffraction. It remains to be seen whether these materials will
also display phase separation between the two types of OO. On one
hand the tendency towards C-type OO increases with smaller
rare-earth radius, but on the other hand the exchange striction at
T$_N$ that appears to promote transformation to the C-type phase
is expected to decrease in magnitude. Nevertheless, it is likely
that the border between the C- and G-type phases as a function of
ionic radius is not a sharply defined line, as suggested by the
phase diagram of Miyasaka et al. \cite{202}, but rather occurs via
a broad phase separated region.
\par We propose a modified RVO$_3$
phase diagram in Fig.\ref{phase} based on diffraction data. For
small rare-earths such as Yb, Ho and Y, complete transformation
from G-type to C-type OO is achieved on cooling through a
first-order transition well below T$_N$. For large rare-earths
such as La, Ce and Pr \cite{266} the orbital ordering remains of
G-type down to at least 5 K. However, for intermediate rare-earths
such as Sm, a region of coexistence between C-type and G-type
phases is present at all temperatures below T$_N$. In summary,
high resolution X-ray diffraction has demonstrated that vanadium
spin ordering in SmVO$_3$ induces magnetic exchange striction that
changes the symmetry of the orbital ordering in part of the
sample. The two types of orbital ordering then coexist down to low
temperature, a situation that is stabilized purely by lattice
strains associated with the difference in unit cell volumes of the
two phases.
\par We thank Dr. I. Margiolaki for experimental assistance at ESRF
and for useful discussions on the X-Ray diffraction experiments.
We are grateful to Prof. D.I. Khomskii for enlightening and
stimulating discussions. We acknowledge financial support by the
European project SCOOTMO RTN (Contract No. HPRN-CT-2002-00293).

\end{document}